\documentclass{article}
\usepackage{amsmath,amssymb}
\usepackage[dvips]{graphicx}

\usepackage{latexsym}

\begin{document}

\pagestyle{plain} 
\setcounter{page}{1}
\setlength{\textheight}{700pt}
\setlength{\topmargin}{-40pt}
\setlength{\headheight}{0pt}
\setlength{\marginparwidth}{-10pt}
\setlength{\textwidth}{20cm}

\title{Separation Number and Generalized Clustering Coefficient in Small World Networks based on String Formalism}
\author{Norihito Toyota   \and Hokkaido Information University, Ebetsu, Nisinopporo 59-2, Japan \and email :toyota@do-johodai.ac.jp }
\date{}
\maketitle

\begin{abstract}
We reformulated  the string formalism given by Aoyama, using an adjacent matrix of a network and introduced a series of generalized clustering coefficients  based on it.   
Furthermore we numerically evaluated Milgram condition proposed by their article in order to explore $q$-$th$ degrees of separation in scale free networks. 
In this article, we apply the reformulation to small world networks and numerically evaluate Milgram condition, especially the separation number of small world networks and its relation to cycle structures are discussed. 
Considering the number of non-zero elements of an adjacent matrix, the average path length and Milgram condition, 
we show that the formalism proposed by us is effective to analyze the six degrees of separation, 
especially effective for analyzing  the relation between the separation number and cycle structures in a network. 
 By this analysis of small world networks, it proves that a sort of power low holds  between $M_n$, which is a key 
quantity in Milgram condition,  and the generalized clustering coefficients. 
This property in small world networks  stands in contrast to that of scale free networks. 
 \end{abstract}
\begin{flushleft}
\textbf{keywords:}
small-world networks, scale free networks, generalized clustering coefficients, six degrees of separation
\end{flushleft}

\section{Introduction}\label{intro}
\hspace{5mm} 
Half a century ago, Milgram has found the phenomenon called "Six degrees of separation"\cite{Milg}  by making a social experiment. 
This experiment  inspired studies of many researchers after that\cite{Albe3},\cite{Newm},    
A series of examinations by him and his coworkers\cite{Milg2},\cite{Milg3}  made a suggestion, 
 which all people in USA are connected through about 6 intermediate acquaintances, more certain.
 Though there are some criticisms in their results, experiments to corroborate their experiments were attempted in various groups 
and theoretical discussions on them also were made\cite{Klein},\cite{Watt4},\cite{Watt3}.   
 At the end of the twenty  century, some breakthroughs have come in the research of network theory 
such as a discovery of small world networks\cite{Watt1},\cite{Watt2} and scale free networks\cite{Albe2} and so on\cite{Doro1},\cite{Doro2}.  
 The understanding of the six degrees of separation deepened through such breakthroughs. 
The understanding of the phenomenon, however, is  insufficient, especially how  cycle structures in a network affect  the six degrees of separation, more generally the separation number are still obscure. 

The first important study of the effect of cycle structures on the separation number in a network has been made by Newman\cite{Newm21}.  
But he considered only the effects of the cycle structures with 3 and 4 nodes, which are triangular and quadrilateral structures.   
We think that it is difficult to pursue the investigation further according to his consideration. 
 Recently Aoyama et al.\cite{Aoyama} developed a method, "string formalism", that could generally analyze the subject. 
 They proposed "Milgram condition" to analyze the the separation number.   
Their evaluation of the separation number, however,  was made with a tree approximation in scale free networks throughout. 
Thus the effect of cycle structures on the separation number are not explored yet.     

We attacked this subject based on the string formalism by fusing it into adjacent matrix description \cite{Toyota3}, \cite{Toyota4},  \cite{Toyota5},  
especially how cycle structures in networks affect the separation number $q$.  
By this, it became really possible to discuss up to just the six degrees of separation, 
while possible up to any degrees of separation in principle.   
This reformation easily make us  extend the usual clustering coefficient\cite{Watt1}, 
which is an index of the number of triangles  in a network, 
to a series of generalized clustering coefficients that give indices  of the number of cycles with any nodes.  
We showed that six degrees of separation has a close relationship with the scale free network with the exponent 3 
by applying this formulation to scale free networks\cite{Toyota6}, \cite{Toyota7}.  
This result is attractive since most of scale free networks in the real world have about the exponent 3.  

In this article, we apply this formulation to small world networks. 
The separation number of small world networks and its relation to cycle structures are discussed. 
By comparing the analyses in this article with the results obtained in scale free networks, 
we show that there are crucial differences in the relation between the separation number and cycle structures in both networks.   
Through the considerations of  the number of non-zero elements of an adjacent matrix, the average path length and Milgram condition, 
we show that the formalism proposed by us is effective to analyze the six degrees of separation, 
especially  for analyzing  the relation between the separation number and cycle structures in a network. 
This  indicates that our formalism gives an appropriate methodology in network analyses.

\section{Reformation of String Formalism based on Adjacent Matrix} 
\hspace{5mm}
In this section we review the formalism given in \cite{Toyota3}, \cite{Toyota4}, \cite{Toyota5}, where the reformulation of the string formalism 
proposed by Aoyama\cite{Aoyama}  by the adjacent matrix is given.  

 We consider a string-like part of a graph with connected $j$ nodes and call it "$j$-$string$" following Aoyama. 
Let $S_j$ be the number of $j$-$string$ and $\bar{S}_j$ be the number of non-degenerate $j$-$string$ on graph. 
The non-degenerate string is defined as the string that does not has any multi-edges and closed cycle structures as a subgraph in a string. 
We, however, consider strings homeomorphic to a circle, called closed strings,  as the non-degenerate string.  
So $\bar{S}_j $ is the total number of the closed strings and the open strings that do not have any closed cycles within themselves.     
It is generally so difficult to calculate $S_j$ and $\bar{S}_j$,  and would be impossible to practically calculate $\bar{S}_j$ 
with $j>7$ at present \cite{Aoyama}.  

By using the reformulation, we can represent the usual clustering coefficient which essentially counts the number of triangles in a network. 
Although there are some definitions of the clustering coefficient \cite{Newm21},\cite{Watt1}, 
we adopt the global clustering coefficient $C_{(3)}$ \cite{Newm21} defined by 
\begin{equation}
C_{(3)}=\frac{6\times \;number \;of \;triangles }{number \;of \;connected \;triplets }=\frac{ 6\Delta_3  }{\bar{S}_3},  
\end{equation}
where $\Delta_3$ is  the number of triangles in a network. 
But we need more indices in order to uncover the effect of general polygon structures in a network. 
From Eq.(1), we can generalize it to $p$-$th$ generalized clustering coefficient $C_{(p)}$ straightforwardly\cite{Toyota3}, \cite{Toyota4}, \cite{Toyota5};  
\begin{equation}
C_{(p)}=\frac{2p\times \;number \;of \;polygons }{number \;of\; connected \;p\mbox{-}plets }=\frac{ 2p\Delta_p  }{\bar{S}_p},
\end{equation}
where $\Delta_p$ is generally the number of polygons with $p$ edges in a network. 

We reformulate the string formalism by utilizing an adjacent matrix $A=(a_{ij})$. 
By doing it, we succeed in this fused formalism to systematically evaluate $\bar{S}_j$ and so $C_{(p)}$. 
Generally the powers, $A^2, A^3, A^4 \cdots$ of $A$ give information as to respecting that a node connects other nodes 
through $2, 3, 4, \cdots$intermediation edges, respectively. 
The matrix elements  of $A^n$  indicates the multiplicity of the connectivity between two nodes, generally. 
So $A^n$ is not suitable for evaluating the number of non-degenerate strings $\bar{S}_j$. 
For resolving the degeneracy of the multiplicity, we introduce new series of matrices $R^n$
which give information as to the connectivity of two nodes, $i_0$ and $i_n$, through $n$ intermediation edges without multiplicity. 
We could find that  $R^n$ with $n>1$  is given by the following formula \cite{Toyota3}, \cite{Toyota4}, \cite{Toyota5}; 
\begin{equation}
[ R^n] _{i_0i_n}=\displaystyle \sum_{i_1,\cdots,i_{n-1}} a_{i_0i_1} a_{i_1i_2}\cdots a_{{i-1},i_{n}} 
\frac{\displaystyle\prod_{i_k,i_j,i_k-i_j>1}^{n}(1-\delta_{i_ki_j})}{(1-\delta_{i_0i_n})}.
\end{equation}
Here the product of the Kronecker delta in the numerator plays role of excluding degeneracies strings or multiplicities   
and the Kronecker in the denominator is needed to keep closed strings, respectively.  
Though Eq.(3) makes one count $\bar{S}_j$ in a unified way, it is unrealistic to directly evaluate the elements of $R^n$ from its expression, 
since this contains multi-loop calculations coming from $\sum$ symbol. 
Even if we expand Eq.(3), we have $2^{n(n-1)/2}$ terms. 
This is 32768 for $n=6$ that is needed to analyze six degrees of separation. 
Thus we can not evaluate $R^n$ within realistic time even in a little large networks.  
Careful expanding of Eq.(3), however, cause drastic cancellation among the terms.      
 We could get some compact expressions for $R^n$ for $n=2\sim6$ in the long run. 
We, however, still need to write down  $R^6$ about a few pages in A4 size.  
The explicit expressions of $R^2\sim R^6$ are described in the references \cite{Toyota3}, \cite{Toyota4}, \cite{Toyota5}.    
 
By using $R^n$, we obtain the expressions for $S_p$ and $C_{(p)}$; 
 \begin{equation}\bar{S}_p=\sum_{i,j} (R^{p-1})_{ij}/2,\end{equation}
\begin{equation}
C_{(p)}=\frac{\mbox{Tr} R^p  }{ \displaystyle \sum_{i,j}R^{p-1}}.  
\end{equation}
As an example, we obtain the following formula for the usual clustering coefficient; 
\begin{equation}
[R^2]_{ij }= [A^2]_{if} - [A^2]_{ii} \delta_{if}, \;\;\; ||A||\equiv\sum_{ij} A_{ij},   
\end{equation}
\begin{equation}\displaystyle 
C_{(3)}=\frac{\mbox{Tr} R^3  }{ \displaystyle \sum_{i,j} (A^2)_{ij}-(A^2)_{ij} \delta_{ij}  }=\frac{ \mbox{Tr} A^3  }{ 
 ||A||-\mbox{Tr} A^2 }.
\end{equation}

\section{Application to $q$-th degrees of separation }
\label{usage}
\subsection{Milgram Condition }
\hspace{5mm}
We analyze general $q$-$th$ dedrees of separation, according to our formalism. 
For $q$-$th$ degrees of separation, Aoyama proposed a condition, so-called Milgram condition 
at network size $N$ \cite{Aoyama}; 
\begin{equation}
M_{q} \equiv \frac{\bar{S}_{q}}{N} \sim O(N),
\end{equation}    
where it is considered that the contribution of strings homeomorphic to a circle can be  ignored at the limit of $N\rightarrow \infty$.  
This condition means that the number of $q$-strings per node is nearly equal to the size of the considering network and gives a boundary whether $q$-$th$  degrees of separation is fulfilled or not. 
This is a natural condition which indicates that a whole network are basically connected each other by $q$ steps.  
As $\bar{S_q}$ grows larger, more exactly $M_q$ grows larger under fixed $N$, $q$-$th$ degrees of separation is easier to be fulfilled. 
$\bar{S_q}$ in Eq.(8) can be calculated from Eq.(4) by using $R^n$. 
We here place the focus on small-world networks.  

\subsection{Application to Small World Networks }
\hspace{5mm}
Milgram condition has been applied to scale free networks in \cite{Toyota6},\cite{Toyota7} already.   
In this article,  we apply Milgram condition to small world networks 
and clarify the difference between the two types of networks.    
It is showm that the string formalism described by this article is available to analyze the relations 
between the separate number and cycle structures through the results.

We construct small world  networks with any rewiring ratio according to Newman-Watts model  \cite{Newm22}. 
For it,  we choose a regular lattice homeomorphic to  one dimension circle $S^1$ with the degree 4  
as a basic network for constructing the networks.     
 Let $\alpha$ be a rewiring parameter and network size is $N=200$.  
Since  the estimation of $R^n$ presents great  computational complexity,
it is difficult to evaluate  $R^n$ for large $n$  within realistic computational time. 
Even in this size, we find that $C_{(3)}$ and the average path length show  characteristic behaviors  
of small world networks.  (Fig. 1 and Fig.4  shown later indicate these facts. )  

We need to estimate $C_{(3)}\sim C_{(6)}$  as indecis that reflect cycle structures in a network 
when we consider six degrees of separation. 
So we defined the following quantity; 
\begin{eqnarray}
X_p &\equiv& \sum_{q=3}^{p} C_{(q)},
\end{eqnarray}
where $q$ is taken from 3, since $C_{(1)}$ and $C_{(2)}$ do not participate in any cycle structures.  
$X_p$ take on the responsibility of information of cycle structures formed by nodes having the number of edges within $p$. 

 \begin{figure}[t]
\begin{center}
\includegraphics[scale=0.9,clip]{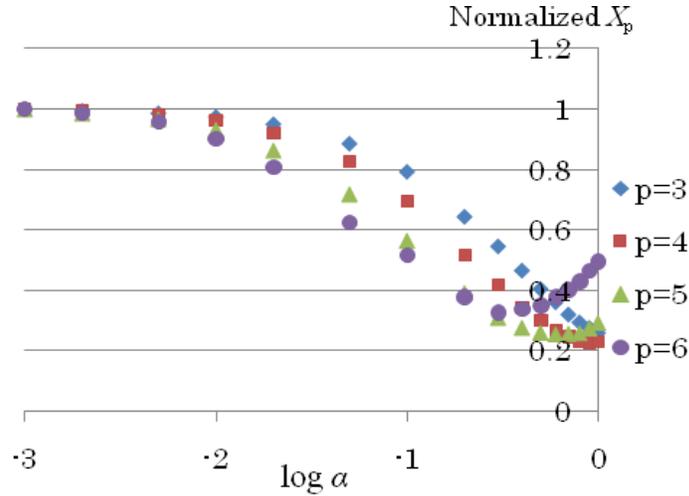} 
\caption{Normalized $Xp$.  }
\end{center}
 \end{figure} 
 
 Fig.1 shows $X_p$-$\log_{10} \alpha$ plot for $p=3\sim6$. 
 The vertical values are normalized by the maximal values of $X_p$ every $p$. 
 A series of data in the top shown by the diamond mark are ones at $p=3$ and date plots in the lower parts 
 represent ones at $p=4, 5, 6$, respectively. 
It may be laid down as a general rule that $X_p$ behaves as the usual clustering coefficient $C_{(3)}$.  
 The networks can be called "generalized small world networks" in that meaning.  
$X_{5,6}$ increase in some degree at $\alpha=0$ where the network is a random network. 
This is because of the finite size effect by starting with rather small size circle in the network construction. 
As $p$ grows larger, the variation in $X_p$ becomes smaller in Fig.1. 
This is naturally understood by noticing that there is no cycles  with larger number than 3 nodes 
in the initial regular network.   
 \begin{figure}[t]
\begin{center}
\includegraphics[scale=0.9,clip]{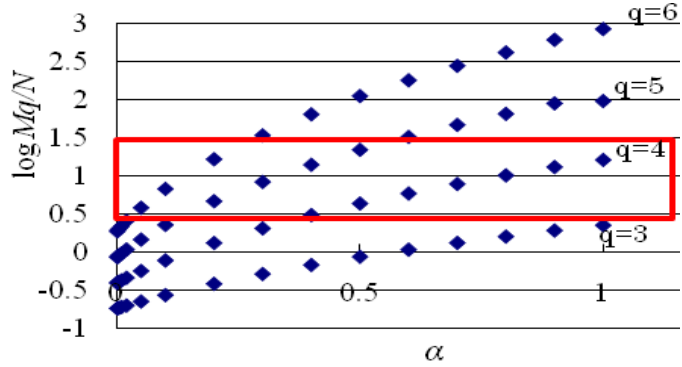} 
\caption{Milgram condition with  $q=3\sim6$.  }
\end{center}
 \end{figure}

 To study the relation between $X_p$ and six degrees of separation, we find the relation between $\alpha$ and $M_p$. 
 $\log M_q/N$ ever  $\alpha$ is shown in Fig.2. 
 Four series of data represent data at $q=3,4,5,6$, respectively.   
The data within the red  rectangle  in Fig.2 mean they satisfy Milgram condition. 
From this, we find that Milgram condition begin to be satisfied just when $X_p$ begin to decrease in Fig.1. 
Thus six degrees of separation is achieved by adding rather a small number of short cuts to the initial regular lattice.    
Though Fig.2 also shows that while four degrees of separation can  be achieve, it is difficult that three degrees of separation can  be achieved even if $\alpha$ comes close to one.   

The consideration of an adjacent matrix also supports that this property is right. 
Let be $r_n$ the ratio of the number of the nonzero elements among all elements in $A^n$. 
$r_n$ give  information about the ratio of nodes connected each other at $n$ steps.  
When considering $q$-$th$ degrees of separation, we need information about nodes connected each other 
within $q$ steps, that is, $n\leq q$. 
We define $T_n$ as the sum of $r_n$;
\begin{equation}\displaystyle 
T_q= \sum_{n=1}^{q} r_n.
\end{equation}
$T_q$-$\alpha$ plot is described in Fig.3 where the data shown "sequence $n$"  represent the data of $T_{q+1}$. 
There are some date with $T_q>1$ in Fig.3,  because there are situations which two nodes are connected each other 
at some different steps.      
When we choose  $T_n= 0.5$ as borderline that means two nodes are connected each other with the possibility of 50 percent including multi-connection, this line corresponds to the center of the region satisfying Milgram condition in Fig.2 ( $\log M_q/N \sim 1$ ).   
  The  situation that two nodes are connected each other with the probability larger than  0.5 
corresponds to the critical point of Milgram condition.

 \begin{figure}[t]
\begin{center}
\includegraphics[scale=0.9,clip]{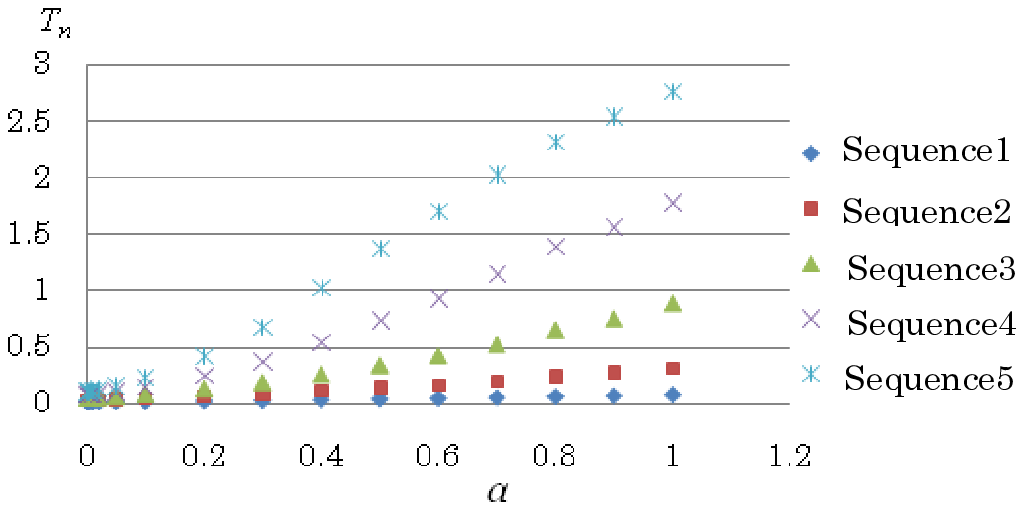} 
\caption{Numbers of non-zero elements  in  adjacency matrices.    }
\vspace{7mm}
\includegraphics[scale=0.9,clip]{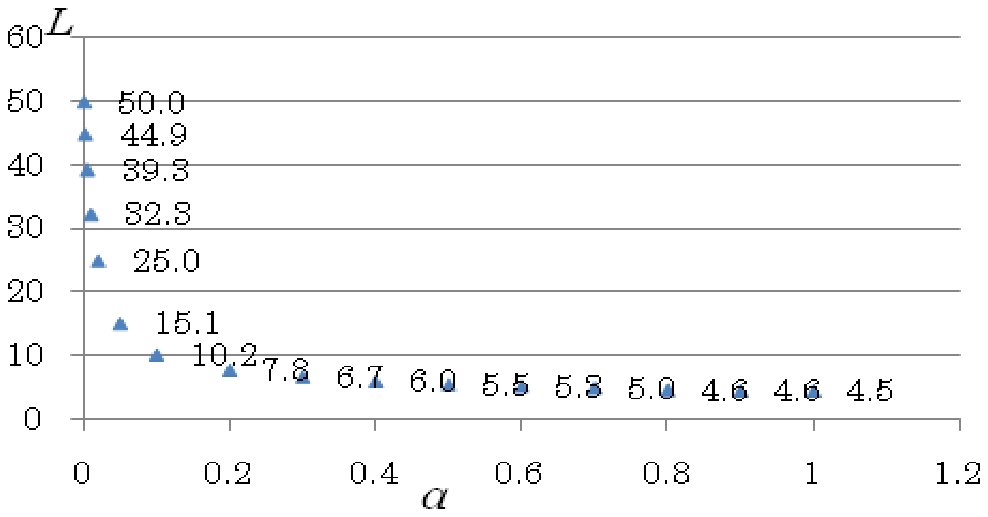} 
\caption{Average path length in small-world networks with  several $\alpha$  }
\end{center}
 \end{figure} 
We explore the relation between the average path length $L$ and $\alpha$.  
The relation is described with numerical data of $L$ in Fig.4. 
Fig.4 shows that $L$ rapidly decreases when $\alpha$ grows a little smaller as usual small world networks.  
Considering $L\sim 10$ conscious of  the separation number almost $\leq 10$, it is consistent with the outcomes referred 
in Fig. 2 and Fig.3.    
So three considerations given from Fig.2,  to Fig.4 almost lead to similar conclusions on the separation number $q=n$.  
$X_p$ among them, however, plays an important role in the discussion on the relation 
between the separation number and cycle structures in a network.  
 This could give a  significant methodology to analyze them.    

We observe that it becomes easier to realize six degrees of separation or $q$ becomes smaller, 
as the (generalized) clustering coefficient grows smaller from Fig.1 and Fig.2.  
 We investigate this a little more quantitatively.  
 Fig.5 is the  graph that represents the relation between  $\log X_n$ and $\log M_n/N$. 
 Each striated data points corresponds to ones at $n=3 \sim 6$. 
We observe that they are almost arranged in straight lines.  
This means the relation between $X_p$ and $M_n$ is given by 
\begin{eqnarray}
M_n &\sim & B_n (X_n)^{-a},  
\end{eqnarray}
where $B_n$ and $a>0$ are some constants that are determined by numerical data. 
This  stands in contrast to the results of scale free networks shown in Fig.6 where the relation between them were 
\begin{eqnarray}
M_n &\sim & \exp( cX_n),    
\end{eqnarray}
  where $c>0$ is a constants that are determined by numerical data\cite{Toyota6}, \cite{Toyota7},. 
 Comparing the signs of the parameters ($a$ and $c$) appearing in Eq.(11) and Eq.(12), 
the response of $M_n$ to $X_n$ is opposite in both the networks.   
While it becomes difficult for small world networks to satisfy Milgram condition, 
it  becomes easy for scale free networks to satisfy the condition, 
as the generalized clustering coefficients becomes large.   
  When there are many cycle structures in small word networks, 
released information from a node tend to turn round and round on same cycles 
so that the propagation of the information are seriously  obstructed by cycle structures.  
Contrary to small world networks, cycle structures  produce some shortcuts 
to make the propagation of information smooth in scale free networks, 
which are almost tree structures.   
Thus  the separation number is not related to cycle structures in the same rule in all networks.  
The effect of cycle structures on $q$-$th$ degrees of separation in a network strongly depends on 
the network topology used basically in constructing the network.   

Furthermore the relation between $\log X_n$ and $\log M_n/N$ lies along a line under fixed $n$ 
and the gradient and the intercept  on the y-axis of the line undergo a change with $n$ 
in small world networks as shown in Fig.5.  
This means that the way of realization of $q$-$th$ degrees of separation 
depends on network topologies and is different in every $n$ as cycle structures are changing\cite{Toyota7}.  
 In scale free networks where the degree distribution $P(k)$ is given by $P(k) \sim k^{-\gamma}$
, however,   the relation between $\log X_n$ and $\log M_n/N$ are universal  to $n$ 
when changing the network topologies by changing $\gamma$.  
A common straight line appears in  $\log X_n$ vs.  $ M_n/N$ graph of  under diverse $n$.   
So the way of realization of $q$-$th$ degrees of separation are common to every $n$ 
in scale free networks.

 \begin{figure}[t]
\begin{center}
\includegraphics[scale=0.9,clip]{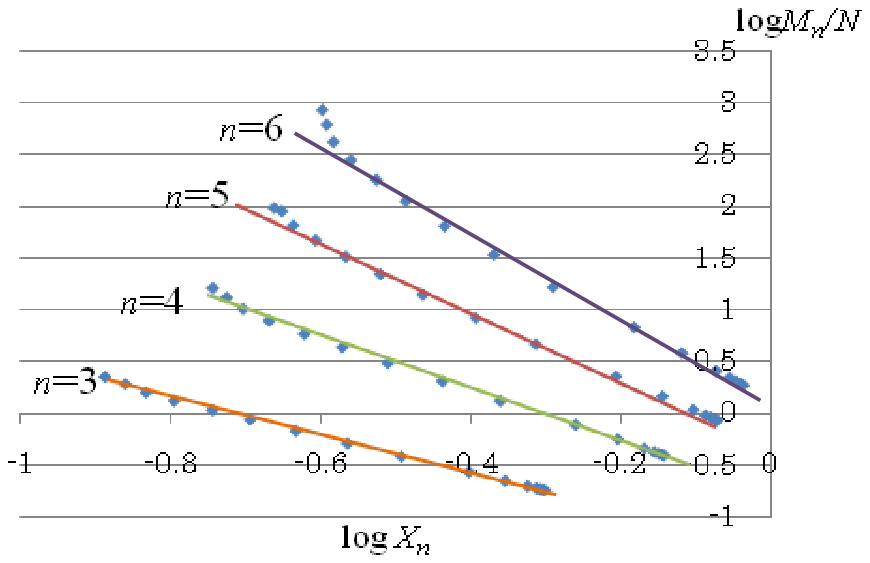} 
\caption{Sum of $C_{(p)}$ v.s.$ \log M_q/N$ in small world networks for every separation number $n$.}
\vspace{10mm}
\includegraphics[scale=0.9,clip]{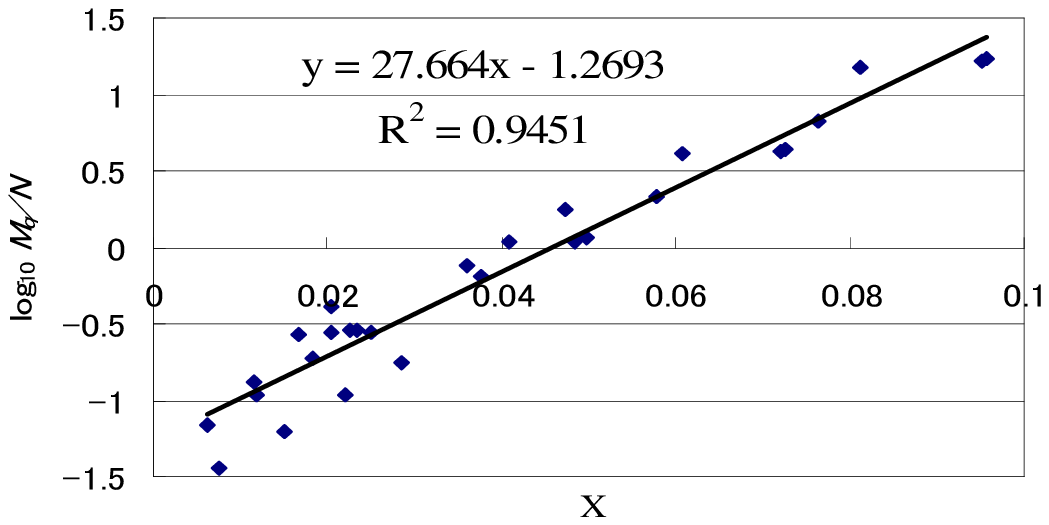}
\caption{ Separation number $q$ v.s.$ \log M_q/N$ in scale free networks with  several $\alpha$  }
\end{center}
 \end{figure}

\section{SUMMARY }
\hspace{5mm}
In this article, we investigate six degrees of separation in small world networks by using reformulation of the string formalism based on an adjacent matrix. 
This reformulation makes it possible for us to systematically evaluate the generalized clustering coefficient.     
Moreover we attacked the problem of general $q$-$th$ degrees of separation based on a series of the generalized clustering coefficients,  especially how cycle structures on networks, whose information is charged with generalized clustering coefficients, affect the separation number $q$. 
Our previous studies support that this reformulation reconstructs the already known properties in scale free networks and 
random networks\cite{Erdos1} with the Poisson distribution in the degree distribution,  
and the formalism is  reliable to analyze properties of networks\cite{Toyota3}, \cite{Toyota4}.    
  The analyses by this formalism also uncovered the relation between the exponent $\gamma$ and six degrees of separation.  
  Especially, we found an interesting  fact that the scale free network with $\gamma= 3.0$, which many real-world networks have about this value of exponent, is closed to six degrees of separation\cite{Toyota5}, \cite{Toyota6}, \cite{Toyota7}. 

As the result of this article, the general clustering coefficients behave like the usual clustering coefficient 
 in small world networks.    
By considering Milgram condition for the separation number,  we find that it rapidly gets  easier to satisfy the condition 
by adding only a few edges to a one dimensional regular lattice homeomorphic to a circle. 
This aspect is similar to one of the average path length and is also supported from the perspective of an adjacent matrix. 
Thus the studies of Milgram condition, an average path length and an adjacent matrix give almost same information in a result. 
The string reformulation developed by us, however,  can only  evoke discussion in connection with cycle structures.  
It is a main assertion that the string formalism based on an adjacent matrix carries great significance in this article.  

By this analysis, it also proves that a sort of power low holds  between $M_n$ and the generalized clustering coefficients. 
This property in small world networks  stands in contrast to that of scale free networks. 
In small world networks, cycle structures operates as  resistance to the propagation of information and the separation number rather decreases  when there is not any cycles.  
This means that the effect of cycle structures  on the separation number  strongly depends on the construction method of networks or the basic properties of network.  

To  give a compact and systematic expression of $R^n$ for large $n$ in order to study beyond six degrees of separation is a  future research topic. 
It should be studied that the main results given by this article are also confirmed for larger size networks.  
Anyway extensive calculations and so highly efficient computer are needed to accomplish them. 
It is, however, confirmed by us that the essential behaviors do not depend on the network size  up to a point in scale free networks. 
This fact is a natural  consequence from the definition of the concept of " scale free".




\begin{thebibliography}{99}
\bibitem{Milg}S. Milgram, "The small world problem", Psychology Today 2, 60-67 (1967)
\bibitem{Albe3}A.-L.Barabasi and R.Albert, "edgeed: The New Science of Networks",  Perseus Books Group (2002) 
edgeed: How Everything Is Connected to Everything Else and What It Means for Business, Science, and Everyday Life
Plume ; ISBN: 0452284392 ; (2003/04/29)
\bibitem{Newm}M.E.J. Newman, A.-L.Barabasi  and D. J. Watts, "The Structure and Dynamics of Networks", Princeton Univ. Press, 2006 
\bibitem{Milg2}J. Travers and S. Milgram, "An Experimental Study of the Small World Problem", Sociometry 32, 425 (1969) 
\bibitem{Milg3} C. Korte and S. Milgram, "Acquaintance links between White and Negro populations: Application of the small world method", Journal of Personality and social Psychology 15 (2), pp.101-108   (1970) 
\bibitem{Klein} J.S. Kleinfield, "The small world problem", Society 39(2) pp.61-66(2002):
"COULD IT BE A BIG WORLD? ", http://www.uaf.edu/northern/big$ \_$world.html 
\bibitem{Watt4}D. J. Watts et al., Small World Project-Columbia University. 
http://small world.columbia.edu/
\bibitem{Watt3}P.S.Dodds, R.Muhamad and D.J. Watts, "An Experimental Study of Research in Global \\Social Networks", Science 301, pp.827-829:  \\http://small world.columbia.edu/images/dodds2003pa.pdf (2003)
\bibitem{Watt1}D. J. Watts　and S. H. Strogatz, "Collective dynamics of 'small-world' networks",　Nature,393, 440-442(1998)
\bibitem{Watt2}D. J. Watts, "Six degree-- The science of a connected age", W.W. Norton and Company, New York (2003)

\bibitem{Albe2}A.-L.Barabasi and R.Albert, "Emergence of scaling in random networks", Science, 286, 509-512(1999)
\bibitem{Albe1}R.Albert and A-.L. Barabasi, "Statistical Mechanics of complex networks",Rev. Mod. Phys. 74, 47-97(2002)

 \bibitem{Doro1}S. N. Dorogovtsev, A.V. Goltsev and J.F.F. Mendes, "Pseudo fractal scale-free web", Phys. Rev. E.65, 066122(2002) 
\bibitem{Doro2}S. N. Dorogovtsev and J.F.F. Mendes, "Evolution of Networka", Oxford Univ. Press, Oxford(2003)

\bibitem{Newm21}M.E.J.Newman,"Ego-centered networks and the ripple effect or why all your friends are wired", Social Networks 25 (2003) p.83;arXiv. cond-mat/0111070
\bibitem{Aoyama} H. Aoyama, "Six degrees of separation; some caluculation", SGC library65, " Introduction to Network Science", (2008) in Japanese;
H, Aoyama, Y.Fujiwara, H, Ietomi, Y. Ikeda and W.Soma "EconoPhysics",Kyouritu Shuppan 2008 n Japanese;


\bibitem{Toyota3} N. Toyota, IEICE Thecnical Report, "String Formalism for $p$-Clustering Coefficient-Toward Six Degrees of Separations",NLP2009-49(2009) in Japanese. 
\bibitem{Toyota4} N. Toyota, " $p$-th Clustering coefficients $C_{p}$ and Adjacent Matrix for Networks: Formulation based on String", arXiv:0912.2807
\bibitem{Toyota5}N. Toyota and T. Sakamoto, "$p$-$th$ degrees of separation in the string-adjacent matrix formalism",  sixth symposium of network ecology, 2009 Dec. in Japanese
 \bibitem{Toyota6}N. Toyota and T. Sakamoto, "Six degrees of separation and the generalized clustering coefficients in scale free networks ",  42-th lecture meeting of the Society of Instrument and Control Engineers in Hokkaido, 2010.Feb. in Japanese
\bibitem{Toyota7} Norihito Toyota, “p-th Clustering coefficients and q-th degrees of separation based on String-Adjacent Formulation”，Preprint arXiv:1002.3431
 
 \bibitem{Newm22}M.E.J. Newman and D. J. Watts, "Renormalization group analysis of the small-world network model”, Phys.Lett.A263,  341-346(1999) 
\bibitem{Erdos1}P. Erdos and A. Renyi," On random graphs I", Publicationes Mathematicae Debrecen6, 290-297, 1959

\bibitem{Lind}P.G.Lind, M.C.Gonzalez and H.J.Hermann, "Cycles and clustering in bipartite networks", Phys.Rev.E 72,056127 (2005)
\bibitem{Zhang} P.Zhang, J.Wang, X.Li, M.Li, Z.Di and Y.Fan,"Clustering coefficient and community structure of bipartite networks", Physica A, 387,

\end{thebibliography}

\end{document}